\documentclass[12pt,a4paper]{article}
 \usepackage{amsmath,amssymb,graphicx,epsfig,cite,color}

 \usepackage[left=2.7cm,right=2.7cm,top=2.5cm,bottom=2.5cm]{geometry}
 \usepackage{subfigure}
 \usepackage[T1]{fontenc}
 	\usepackage{float}   
 	\usepackage{array} 
 	\usepackage{makecell}
 	\usepackage{multirow}
 	\usepackage[table]{xcolor}
 	\usepackage{booktabs}
 	\usepackage{array}
 	\usepackage{colortbl}
 	
 	\definecolor{tabcolor1}{rgb}{.105,.410,.113}
 	\definecolor{tabcolor2}{rgb}{.425,.130,.303}
 	\definecolor{tabcolor3}{rgb}{1.00,.0,.0}
 	
 	\usepackage{xcolor,colortbl}

 	\usepackage{graphicx}
 	\usepackage{lscape}
 	\usepackage{appendix}
 	\usepackage{multirow}
 	\definecolor{DarkPastelGreen}{rgb}{0.01,0.69,0.28}
 	\definecolor{amethyst}{rgb}{0.6, 0.4, 0.8}
 	\definecolor{darkcyan}{rgb}{0.0, 0.55, 0.55}
 	\definecolor{lightcyan}{rgb}{0.88, 1.0, 1.0}
 	\definecolor{antiquefuchsia}{rgb}{0.57, 0.36, 0.51}
 	\definecolor{brightube}{rgb}{0.82, 0.62, 0.91}
 	\definecolor{brilliantlavender}{rgb}{0.96, 0.73, 1.0}
 	\definecolor{bazaar}{rgb}{0.6, 0.47, 0.48}
 	\usepackage[colorlinks=true,linkcolor=amethyst,citecolor=brightube,urlcolor  = antiquefuchsia]{hyperref}%
 	\allowdisplaybreaks[1]
 	
\begin{document}




	\title{
		{\Large
			{\bf
				\color{amethyst}{Exploring fully-heavy tetraquarks through the CGAN framework: Mass and width}
			}
		}
	}

			\author{\vspace{1cm}\\
			{\small
				M.~Malekhosseini$^{1}$	,
				S.~Rostami$^{2}$,
				A. R. Olamaei$^{2,3}$,
				K. Azizi$^{1,4}$
				\thanks{Corresponding author, E-mail: kazem.azizi@ut.ac.ir}
			}
			\\
			{\small $^1$ Department of Physics, University of Tehran, North Karegar Avenue, Tehran 14395-547, Iran}\\
			{\small$^2$  School of Physics, Institute for Research in Fundamental Sciences (IPM),}\\
			{\small  P. O. Box 19395-5531, Tehran, Iran}\\
			{\small $^3$ Department of Physics, Jahrom University, Jahrom, P.  O.  Box 74137-66171, Iran}\\
			{\small $^4$ Department of Physics,  Dogus  University, Dudullu-\"Umraniye, 34775 Istanbul, T\"urkiye}\\
		} 
		
		\date{}

		\begin{titlepage}
			\maketitle
			\thispagestyle{empty}
			
			\begin{abstract}
			Fully-heavy tetraquark states, $QQ\bar{Q}\bar{Q} (Q=c,b)$, have garnered significant attention both experimentally and theoretically, due to their unique properties and potential to provide new insights into Quantum Chromodynamics (QCD). In this study, we employ Conditional Generative Adversarial Networks (CGANs) to predict the masses and decay widths of fully-heavy tetraquarks. To deepen our understanding of heavy multiquark structures, we prepare datasets based on two distinct approaches and train the CGAN model using both. The CGAN framework allows us to capture the complex relationships between input features, such as quark content, quantum numbers, and Clebsch-Gordan coefficients, and output properties, including mass and decay width. Our predictions, based on the CGAN framework, are consistent with existing data. By combining fundamental knowledge of QCD with advanced machine learning techniques, this work represents a significant step forward in the theoretical understanding of fully-heavy tetraquark states. Our CGAN approach has the potential to become a strong contender for future studies in heavy tetraquark systems, complementing existing theoretical models to deliver more precise results. Additionally, our findings could assist in the search for fully-heavy tetraquark systems in future experiments.
			\end{abstract}

		\end{titlepage}
		\section{Introduction}
		
Exotic hadrons are particles that extend beyond the traditional hadron bound states, 
which includes three-quark baryons and quark-antiquark mesons. Tetraquarks, 
a type of exotic hadron composed of two quarks and two antiquarks, 
present a more complex structure that challenges the conventional quark model 
\cite{Gell-Mann:1964ewy, ParticleDataGroup:1986kuw, ParticleDataGroup:2022pth, ParticleDataGroup:2024cfk}.	
The concept of multi-quark states was first introduced by Gell-Mann \cite{Gell-Mann:1964ewy}
and Zweig \cite{Zweig:1964jf}. Over the past fifty years, this topic has evolved into a captivating research field, 
providing deeper insights into the QCD theory of the strong interaction		
\cite{Jaffe:1976ig, Jaffe:1976ih, Ader:1981db, Pepin:1996id, Brink:1998as}.
However, it is only in recent decades that experimental advances have enabled the detection of possible tetraquark candidates, 
such as the $X(3872)$ by the Belle collaboration \cite{Belle:2003nnu}.		
This eventually led to the discovery of several tetraquark states, such as the 
$Z_{c}(3900)$, $Z_{c}(4020)$, $Z_{cs}(3985)^{-}$ and $Z_{cs}(4220)^{+}$ \cite{Ortega:2023pmr, BESIII:2020qkh, LHCb:2021uow}.		
The discovery of these exotic particles has opened up new avenues of research in hadronic physics. 
In particular, fully-heavy tetraquarks, composed solely of heavy quarks, charm (\( c \)) and bottom (\( b \)),
have attracted significant attention in recent years.			
From an experimental perspective, it is believed that identifying 
fully-heavy tetraquark states among the already observed ones should be straightforward, as their masses are expected 
to differ significantly from those of conventional heavy mesons. 
From the theoretical perspective, they are expected to exhibit greater stability than their lighter counterparts. 
Regarding the interaction between heavy quarks, chiral symmetry is explicitly violated, 
meaning that meson-exchange forces cannot play a role in a fully-heavy tetraquark system. 			
This situation would encourage the formation of true tetraquark configurations 
instead of loosely bound hadronic molecules which 
typically lead to instability in lighter tetraquark systems \cite{Liu:2019zuc, Alcaraz-Pelegrina:2022fsi}.			
These distinctive properties of fully-heavy tetraquarks make them particularly interesting. 
As a result, they have the potential to uncover new insights into QCD, 
especially in the non-perturbative regime of the strong interaction. 

Not long ago, the LHCb collaboration announced the discovery of $X(6900)$, 
with the quark content of $c\bar{c}c\bar{c}$. 
They measured the mass and width of the narrow $X(6900)$ structure, 
assuming a Breit–Wigner lineshape \cite{LHCb:2020bwg}.			
The X(6900) structure, initially observed by LHCb, 
has also been confirmed by the ATLAS and CMS collaborations	\cite{Xu:2022rnl, CMS:2023owd}.	
Furthermore, a search was conducted by the LHCb collaboration to investigate 
a possible exotic state composed of two $b$	quarks and two $\bar{b}$ quarks, 
$X_{b\bar{b}b\bar{b}}$, but no significant excess was found \cite{LHCb:2018uwm}.		

Alongside experimental efforts to explore possible fully-heavy tetraquarks, 
theoretical research in this area has also been active. 
The study of heavy multiquark states dates back to the early stages of exotic hadron research 
\cite{Iwasaki:1976cn, Chao:1980dv, Badalian:1985es}. 
Recent discoveries of exotic states by experimental collaborations have 
further encouraged theoretical groups to investigate fully heavy tetraquarks. 
As a result, numerous studies have focused on fully heavy tetraquarks in recent years				
\cite{Berezhnoy:2011xn, Karliner:2016zzc, Agaev:2023wua, Agaev:2023gaq, Agaev:2024uza, Chen:2016jxd, Gordillo:2020sgc, Yang:2021hrb, Deng:2020iqw, Wu:2024hrv, An:2022qpt, Richard:2018yrm, Richard:2017vry, Vijande:2009kj, Anwar:2017toa, Wang:2017jtz, Esposito:2018cwh, Hughes:2017xie, Agaev:2023tzi, Agaev:2022ast, Ali:2017jda, Chen:2022asf}.
Some studies predict the existence of several fully-heavy tetraquarks, 
both with symmetric and asymmetric quark contents, based on the QCD sum rule method. 
Their masses and decay widths have been accurately estimated using this approach \cite{Agaev:2023wua, Agaev:2023gaq, Agaev:2024uza}.
For instance, Ref. \cite{Agaev:2023wua},	
performed a detailed analysis of the scalar diquark-antidiquark state of $X_{4c}$ 
using the QCD sum rule and calculated its mass and decay width. 
Their findings appear to be consistent with experimental results. 
Also, with the same method, the mass and width of the fully charmed and beauty tetraquarks 
$T_{4c}$ and $T_{4b}$ were predicted \cite{Agaev:2023gaq}.
Moreover, the QCD sum rule method is exploited to investigate 
the scalar tetraquarks $T_b$ and $T_c$ with asymmetric contents 
$bb\bar{b}\bar{c}$ and $cc\bar{c}\bar{b}$ and their masses and widths were calculated \cite{Agaev:2024uza}.
A moment QCD sum rule method, enhanced by fundamental inequalities, 
was developed to investigate the existence of exotic doubly hidden-charm and bottom tetraquark states composed of four heavy quarks.
Using this method, the mass spectra of these tetraquarks were also obtained \cite{Chen:2016jxd}.
A diffusion Monte Carlo technique was employed to solve the many-body Schrödinger equation 
for fully-heavy tetraquark systems. This analysis offered precise insights into 
the mass spectrum of the all-heavy tetraquark ground states	\cite{Gordillo:2020sgc}.
Ref. \cite{Deng:2020iqw} employed various models, such as the color-magnetic interaction model, 
the traditional constituent quark model, and the multiquark color flux-tube model
to systematically examine the properties of the states $[Q_1Q_2][\bar{Q_3}\bar{Q_4}]$ with $Q_i=c,b$. 
Their findings indicate that the Coulomb interaction plays a crucial role in the dynamical model calculations for heavy hadrons, 
leading to the conclusion that no bound states of the form $[Q_1Q_2][\bar{Q_3}\bar{Q_4}]$ can be observed in the dynamical models.
Fully-heavy tetraquark states, $QQ\bar{Q}\bar{Q} $ 
were systematically studied using a non-relativistic quark model based on 
lattice-QCD findings on the two body $Q\bar{Q}$ interaction \cite{Yang:2021hrb}.
In Ref. \cite{Wu:2024hrv}, the quark potential model was used to compute 
the mass spectrum of the S-wave fully heavy tetraquarks with various flavors. 
The authors applied the Gaussian expansion method to solve the four-body Schrödinger equation and 
employed the complex scaling method to identify resonant states. 
Also, Ref. \cite{An:2022qpt} studied all possible configurations 
for the ground states of fully heavy tetraquarks using the constituent quark model. 
They examined spectroscopy behavior of fully heavy tetraquarks such as binding energy, 
specific wave function, magnetic moment, internal mass contributions and other related characteristics. 

While theoretical approaches like QCD sum rules and lattice-QCD have been employed 
to predict the properties of fully heavy tetraquarks, including mass spectra and decay widths, 
the application of modern machine learning (ML) techniques, 
which have revolutionized many areas of physics, remains nearly absent in this context. In fact, we have recently developed a CGAN framework to predict the masses of some fully-heavy tetraquark states
\cite{Rostami:2025sff}.		

In this work, we aim to expand on our recent study by using the CGAN framework
to predict both the masses and decay widths of fully-heavy tetraquarks, 
employing two distinct approaches for preparing the dataset.
This analysis is specifically devoted to  fully-heavy tetraquarks and
will cover a larger number of these challenging multi-heavy quarks bound states. 
Indeed, our approach not only seeks to predict  the masses and decay widths of fully-heavy tetraquarks accurately, 
but also helps advance the use of the ML techniques in the hadron physics, 
opening new pathways to study the complex behavior of more exotic states.

The paper is organized as follows. In section \ref{CGAN}, 
a brief introduction to the GAN and CGAN models is provided,
along with their applications in particle physics research.
In section \ref{Res}, we present the numerical results of our analysis 
and compare them with existing theoretical predictions.
Lastly, Section \ref{SC} is dedicated to our concluding remarks.

\section{GAN Applications in Particle Physics: An Overview}\label{CGAN}
Recently, deep learning technology has emerged as the leading computational method in ML, 
delivering outstanding performance on a wide range of complex cognitive tasks, 
often rivaling or even exceeding human capabilities \cite{Roberts:2021fes}.
One notable advancement within deep learning is the development of deep generative models, 
which focus on learning the underlying distribution of data and generating new, realistic instances, 
enabling applications in various fields, including high energy physics (HEP)
\cite{Hashemi:2023rgo, Paganini:2017dwg, Salamani:2018uka, Paganini:2017hrr}.
Generative Adversarial Networks (GANs), a novel approach for training deep generative models, 
frame the task as a competitive interaction between two networks: a generator, \( G \) , and a discriminator,  \( D \), 
each trying to outperform the other \cite{Goodfellow:2014upx}.
The generator produces new data samples, \( G(z) \), by transforming random noise, \( z \), 
while the discriminator assesses their realism by contrasting them with real data, \( x \).
The objective function of a GAN can be expressed as,
\[
\min_G \max_D V(D, G) = \mathbb{E}_{x \sim p_{\text{data}}(x)}[\log D(x)] + \mathbb{E}_{z \sim p_z(z)}[\log(1 - D(G(z)))],
\]
where \( p_{\text{data}}(x) \) is the distribution of real data, 
and \( p_z(z) \) is the distribution of the noise.
The generator and discriminator are trained together, with the generator refining its ability to produce convincing fake samples, 
while the discriminator becomes more skilled at identifying them as fake.
CGANs build upon GANs by including extra information, 
like class labels or other conditional data  \( c \), into the generation process \cite{Mirza:2014dfp}.
In a CGAN, both the generator and discriminator are provided with this conditional information, 
enabling the network to generate data samples that are not only realistic 
but also tailored to the given conditions. The objective function of a CGAN is modified as follows,
\[
\min_G \max_D V(D, G) = \mathbb{E}_{x \sim p_{\text{data}}(x)}[\log D(x|c)] + \mathbb{E}_{z \sim p_z(z)}[\log(1 - D(G(z|c)))],
\]
where \( c \) represents the conditional information.
This makes CGANs particularly useful for tasks where the generated data must adhere 
to specific constraints or properties.

In particle physics, the GAN architecture was first applied to produce jet images—2D representations of
energy depositions from particles interacting with a calorimeter. 
This study is considered one of the earliest successful uses of GANs in the physical sciences
\cite{deOliveira:2017pjk}. 
To lower the computational cost of simulating electromagnetic showers, 
GANs were applied to directly generate the component read-outs from electromagnetic calorimeters. 
This method significantly reduced the need for full-scale simulations while maintaining accuracy of the data \cite{Paganini:2017hrr}.
In HEP, the output of certain detectors, such as calorimeters, can be interpreted as images. 
Therefore, techniques used in image recognition can be applied to analyze the output from these detectors. 
Ref. \cite{Carminati:2018khv} represents the first application of three-dimensional convolutional GANs in HEP simulation. 
In this study, three-dimensional images of particles depositing energy in high-granularity calorimeters were generated. 
The results show that GAN models are promising candidates for fast simulations of high-granularity detectors, 
which are expected to be used in next-generation accelerators.
Besides, a quantum GAN-based approach was developed for anomaly detection in particle physics, 
specifically targeting events that can not be explained by the Standard Model. 
The method offers advantages over classical techniques, 
including the ability to work with less training data and improved modeling power. 
The model's potential is demonstrated through simulations and real quantum processor experiments, 
showing it could be a valuable tool for identifying new, 
unknown phenomena in the data produced by next-generation particle colliders \cite{Bermot:2023kvh}.
In a related study, GAN models were employed to explore the possibility of generating large quantities of analysis-specific 
simulated LHC events. The goal was to reduce the computational cost of generating these events 
by training the GANs to directly generate high-level features essential for specific physics analyses, 
such as muon four-momenta in $ Z \to \mu \mu $ events.
This approach indicates the potential of GANs to significantly streamline 
simulations in HEP experiments, making them more efficient and cost-effective \cite{Hashemi:2019fkn}.
Additionally, for the first time, we have employed the CGAN framework to augment the limited dataset of mesons and
predict the mass and width of several challenging mesons \cite{Rostami:2025sff}.
The ability of these networks to model complex distributions and 
generate high-dimensional data makes them invaluable tools in both theoretical and experimental physics.

In this work, we employ the CGAN framework to predict the mass and width of fully-heavy tetraquarks.
We carefully extracted the dataset containing both ordinary and exotic mesons, 
as well as fully-heavy tetraquarks, categorized according to two distinct approaches, 
which were thoroughly explained in our recent study \cite{Malekhosseini:2024eot}.
In the first approach (A1), the data structure is categorized based on the quark content and quantum numbers of the mesons. 
In the second approach (A2), the Clebsch-Gordan coefficients are used in the linear combinations of $q\bar{q}$ sets to modify the dataset.
Thus, the conditional information provided to the CGAN includes the quark content, 
Clebsch-Gordan coefficients and relevant quantum numbers of mesons and fully heavy tetraquarks in both approaches.
In fact, by training the CGAN model on a dataset of known ordinary and exotic mesons' masses and widths, 
the network learns to generate accurate predictions for mass and width of new, unseen fully-heavy tetraquarks.  
The use of CGANs allows us to incorporate specific physical constraints into the generation process, 
resulting in more reliable and physically meaningful predictions.

\section{Results}\label{Res}
In this section, we present the numerical results for the mass and width of fully-heavy tetraquarks, 
predicted using our CGAN model under two distinct approaches (A1 and A2). 
These results are then compared with available experimental data and existing theoretical predictions.

\subsection{$ cc\bar{c}\bar{c} $ and $ bb\bar{b}\bar{b} $ states}  
First we examine the $ cc\bar{c}\bar{c} $ and $ bb\bar{b}\bar{b} $ states.
The LHCb Collaboration observed a di-$J/\psi$ resonance near 6.9 GeV \cite{LHCb:2020bwg}, 
which was widely interpreted in the literature as a fully charm tetraquark. 
This resonance was subsequently confirmed by the ATLAS and CMS Collaborations \cite{Xu:2022rnl, CMS:2023owd}.
The discovery of this new state, composed solely of charm quarks, opened a fresh avenue for exploring the physics of exotic hadrons.
Tables \ref{tab:ccccmass} and \ref{tab:ccccwidth} present the mass and full width predictions for the fully charmed
tetraquark, which were obtained using the CGAN model, based on A1 and A2, alongside a comparison with the experimental data.
As shown in Table \ref{tab:ccccmass}, 
for both A1 and A2, the predicted masses vary across different quantum numbers ($J^{PC}$).
It can be observed that the masses predicted by A2 are generally slightly higher than those from A1 for various quantum numbers. 
For instance, in the $0^{++}$ state, the mass predicted by A2 is $6941 \pm{159} \,~\text{MeV}$, compared to $6878 \pm{154}\,~\text{MeV}$ for A1. 
Despite these differences, both approaches yield results in reasonable agreement with the experimental mass of 
$6763 \pm{681} \,~\text{MeV}$, although the experimental uncertainty is significant.
Following the comparison of predicted masses, we now turn to the analysis of the predicted widths for 
the fully charmed tetraquark, as shown in Table \ref{tab:ccccwidth}.
For certain states, such as $0^{++}$, $2^{++}$, $2^{-+}$ and $2^{+-}$,
our results for both A1 and A2, are compatible with the experimental data.
However, some discrepancies arise for states like $1^{++}$. Despite this, 
the predictions from both A1 and A2 generally fall within a comparable range to the experimental measurement.
This suggests that the CGAN models provide a reasonable estimate for the full width of the $ cc\bar{c}\bar{c} $ state.

As previous studies have shown, in the fully bottomed sector, 
the $ bb\bar{b}\bar{b} $ state has been investigated by LHCb \cite{LHCb:2018uwm} 
and the CMS collaborations \cite{CMS:2016liw, CMS:2020qwa}, 
though no conclusive evidence has been observed so far.
However, the theoretical models have provided valuable insights into the $ bb\bar{b}\bar{b} $ system.
The mass and full width predictions for this challenging fully-heavy tetraquark, based on the CGAN model using A1 and A2, 
are compared with the estimates from these theoretical models in Tables \ref{tab:bbbbmass} and \ref{tab:bbbbwidth}, respectively.
According to Table \ref{tab:bbbbmass}, the CGAN-predicted masses for both A1 and A2 vary across different quantum numbers ($J^{PC}$),
but they remain within a comparable range.
For instance, in $0^{++}$ state, the mass predicted by A1 is $19096 \pm{394} \,~\text{MeV}$, 
while A2 predicts a slightly higher value of $19250 \pm{357} \,~\text{MeV}$. 
When compared with other theoretical models, the predictions of the CGAN model exhibit acceptable agreement.
For example, the QCD sum rule methods, the nonrelativistic constituent quark models, the Monte Carlo method and 
the diquark model offer values in the vicinity of $ 18130 \,~\text{MeV}$ to $ 19352 \,~\text{MeV}$  for the $0^{++}$ state. 
The pattern of consistency is observed for other quantum states ($J^{PC}$), 
where the CGAN model predictions, both for A1 and A2, align with the predictions from other models.
These results suggest that the CGAN model provides reasonable mass predictions for the fully bottomed tetraquark, 
with values that are competitive with well-established theoretical models.
The next step is to compare the CGAN predictions, based on  A1 and A2, 
for the full width of the $bb\bar{b}\bar{b}$ state with the theoretical models presented in Table \ref{tab:bbbbwidth}.
We estimated the full width for various quantum states using the CGAN, 
but the lack of experimental data and the limited availability of theoretical results make a precise comparison challenging. 
Obviously, the predicted widths from A1 are generally higher than those predicted by A2, 
and the CGAN-predicted values for $0^{++}$ are consistent with at least two of the theoretical values in the table.

\begin{table} 
	\renewcommand{\arraystretch}{1.5}	   
	\scalebox{0.46}{
		\centering
		\begin{tabular}{crl|cccc|ccc|ccc|ccc}
			\toprule[1.0pt]
			\toprule[0.5pt]
			\multicolumn{3}{c}{Systems} & \multicolumn{13}{c}{$cc\bar{c}\bar{c}$} \\
			\multicolumn{3}{l}{$J^{PC}$} & \multicolumn{2}{c}{ $0^{++}$} & $1^{++}$ & \multicolumn{1}{c}{$2^{++}$} & $0^{-+}$ & $1^{-+}$ & \multicolumn{1}{c}{$2^{-+}$} & $0^{+-}$ & $1^{+-}$ & \multicolumn{1}{c}{$2^{+-}$} & $0^{--}$ & $1^{--}$ & $2^{--}$ \\
			\toprule[0.5pt]&&Our result A1  & \multicolumn{2}{c}{$ 6878 \pm 154 $ }& $ 6501\pm {130}$& $ 6230\pm {126}$ &  $ 7217\pm {310}$ & $ 6773\pm {238}$ &$ 6532\pm {180}$ & $ 6056\pm {167}$  &$ 5725\pm {180}$&  $ 5604\pm {170}$ &  $ 6249\pm {248}$& $ 6018\pm {170}$ &$ 5957\pm {200}$  \\ 
			&&Our result A2  & \multicolumn{2}{c}{ $ 6941 \pm 159 $} & $ 6711 \pm 190 $ &  $ 6405 \pm 151 $ & $ 6985 \pm 370 $ & $ 6567 \pm 266 $& $ 6266 \pm 335 $ & $ 7319 \pm 260 $ &$ 7032 \pm 178 $ &$ 6973 \pm 144 $ & $ 7150 \pm 300 $ &$ 6979 \pm 223 $ &$ 6792 \pm 229 $ \\
			\Xcline{1-16}{0.01pt}\multicolumn{3}{c}{\makecell[c]{Experimental Mass Ref. \cite{ParticleDataGroup:2022pth}}} & \multicolumn{13}{c}{$6763 \pm{681}$  MeV} \\
  			\toprule[0.5pt]
			\Xcline{1-16}{0.01pt}
			\multicolumn{2}{c}{\multirow{3}*{\makecell[c]{The QCD sum\\ rule method}}}&Ref. \cite{Wang:2021mma}&6540&-&-&6520&7000&6980&-&-&6470&-&7000&6990&-\\
			\Xcline{3-3}{0.01pt}
			&&Ref. \cite{Wang:2017jtz,Wang:2018poa}&\multicolumn{2}{c}{5990}&-&6090&-&-&-&-&6050&-&-&6110&-\\
			&&Ref. \cite{Agaev:2023wua}&\multicolumn{2}{c}{6570}&-&-&-&-&-&-&-&-&-&-&-\\
			\Xcline{1-16}{0.01pt}
			\multicolumn{2}{c}{\multirow{6}*{\makecell[c]{The nonrelativistic\\ constituent\\quark models}}}&Ref. \cite{An:2022qpt}&6384&6512&-&\multicolumn{1}{c|}{6483}&-&-&-&-&6452&-&-&-&-\\
			\Xcline{3-3}{0.01pt}
			&&\multirow{2}*{Ref. \cite{Wang:2019rdo}}&6377&6425&-&6432&-&-&-&-&6425&-&-&-&-\\
			&&&6371&6483&-&6479&-&-&-&-&6450&-&-&-&-\\
			\Xcline{3-3}{0.01pt}
			&&Ref. \cite{Liu:2019zuc}&6487&6518&-&6524&-&-&-&-&6500&-&-&-&-\\
			\Xcline{3-3}{0.01pt}
			&&Ref. \cite{Zhang:2022qtp}&6500&6411&-&6475&-&-&-&-&6453&-&-&-&-\\
			\Xcline{3-3}{0.01pt}
			\Xcline{1-16}{0.01pt}
			\multicolumn{2}{c}{\makecell[c]{The Bethe-\\ Salpeter equations}}&Ref.\cite{Li:2021ygk}&\multicolumn{2}{c}{6419}&-&6516&-&-&-&-&6456&-&-&-&-\\
			\Xcline{1-16}{0.01pt}
			\multicolumn{2}{c}{\multirow{2}*{\makecell[c]{The relativistic\\ quark model}}}&Ref. \cite{Faustov:2020qfm}&\multicolumn{2}{c}{6190}&-&6367&-&-&-&-&6271&-&-&-&-\\
			\Xcline{3-3}{0.01pt}
			&&Ref. \cite{Lu:2020cns}&6435&6542&-&6543&-&-&-&-&6515&-&-&-&-\\
			\Xcline{1-16}{0.01pt}
			\multicolumn{2}{c}{Monte Carlo method}&Ref. \cite{Gordillo:2020sgc}&\multicolumn{2}{c}{6351}&-&6471&-&-&-&-&6441&-&-&-&-\\
			\Xcline{1-16}{0.01pt}
			\multicolumn{2}{c}{\multirow{2}*{\makecell[c]{
						The diquark model}}}&Ref. \cite{Berezhnoy:2011xn}&\multicolumn{2}{c}{5966}&-&6223&-&-&-&-&6051&-&-&-&-\\
			\Xcline{3-3}{0.01pt}
			&&Ref. \cite{Mutuk:2021hmi}&\multicolumn{2}{c}{6322}&-&6385&-&-&-&-&6354&-&-&-&-\\
			\Xcline{1-16}{0.01pt}
			\multicolumn{2}{c}{\makecell[c]{An effective\\ potential model} }&Ref. \cite{Zhao:2020nwy}&6346&6476&-&6475&-&-&-&-&6441&-&-&-&-\\
			\Xcline{1-16}{0.01pt}
			\multicolumn{2}{c}{\makecell[c]{ Multiquark color\\ flux-tube model}}&Ref. \cite{Deng:2020iqw}&\multicolumn{2}{c}{6407}&-&6486&-&-&-&-&6463&-&-&-&-\\
			\Xcline{1-16}{0.01pt}
			\multicolumn{2}{c}{\multirow{5}*{\makecell[c]{The chromo-\\magnetic model}}}&Ref. \cite{Karliner:2016zzc}&\multicolumn{2}{c}{6192}&-&-&-&-&-&-&-&-&-&-&-\\
			\Xcline{3-3}{0.01pt}
			&&\multirow{2}*{Ref. \cite{Wu:2016vtq}}&6899&7016&-&6956&-&-&-&-&6899&-&-&-&-\\
			&&&6035&6253&-&6194&-&-&-&-&6137&-&-&-&-\\
			\Xcline{3-3}{0.01pt}
			&&Ref. \cite{Weng:2020jao}&6045&6271&-&6287&-&-&-&-&6231&-&-&-&-\\
			\Xcline{3-3}{0.01pt}
			&&Ref. \cite{Zhuang:2021pci}&6034&6254&-&6194&-&-&-&-&6137&-&-&-&-\\
			\toprule[0.5pt]
			\toprule[1.0pt]
	\end{tabular}}
	\caption{ \scriptsize{Our CGAN predictions for the mass of the $ cc\bar{c}\bar{c} $ state (in units of MeV), based on A1 and A2, compared with the
			experimental data \cite{ParticleDataGroup:2022pth} and the theoretical estimates.}}	
	\label{tab:ccccmass}
\end{table}

    \begin{table} 
  	\renewcommand{\arraystretch}{1.5}	
  	\scalebox{0.54}{
  		\centering
  		\begin{tabular}{crl|ccc|ccc|ccc|ccc}
  			\toprule[1.0pt]
  			\toprule[0.5pt]
  			\multicolumn{3}{c}{Systems} & \multicolumn{12}{c}{$cc\bar{c}\bar{c}$} \\
  			\multicolumn{3}{l}{$J^{PC}$} &\multicolumn{1}{c}{ $0^{++}$} & $1^{++}$ & \multicolumn{1}{c}{$2^{++}$} & $0^{-+}$ & $1^{-+}$ & \multicolumn{1}{c}{$2^{-+}$} & $0^{+-}$ & $1^{+-}$ & \multicolumn{1}{c}{$2^{+-}$} & $0^{--}$ & $1^{--}$ & $2^{--}$ \\
  			\toprule[0.5pt]&&Our result A1   &$138\pm {91}$ &$124\pm {23}$ &$68\pm {12}$ & $125\pm {50}$& $68\pm {18}$&$71\pm {14}$ &$159\pm {100}$ & $143\pm {39}$&$82\pm {38}$ & $191\pm {97}$& $106\pm {47}$& $86\pm {27}$ \\ &&Our result A2  &$148\pm {38}$ & $61\pm {20}$ &$60\pm {13}$  &$148\pm {29}$  &$35\pm {10}$&$ 58\pm 10 $  &$169\pm {101}$  & $50\pm {12}$  & $39\pm {12}$ & $164\pm {96}$ & $62\pm {15}$ &$30\pm {8}$  \\
  			\Xcline{1-15}{0.01pt}\multicolumn{3}{c}{\makecell[c]{Experimental Width Ref. \cite{ParticleDataGroup:2022pth}}} & \multicolumn{12}{c}{$153 \pm{29}$  MeV} \\
  						\toprule[0.5pt]\multicolumn{2}{c}{\multirow{3}*{\makecell[c]{The QCD sum\\ rule method}}} &Ref. \cite{Agaev:2023wua}& 110 &- & -& -& -&- &- &- & -&- &- & - \\ 
			\Xcline{3-3}{0.01pt}&& Ref. \cite{Agaev:2023gaq}& 128 &-  & -& -& -&- &- &- & -&- &- & - \\
		\Xcline{3-3}{0.01pt}&& Ref. \cite{Yang:2024guo}&70.18 &-  & -& -& -&- &- &- & -&- &- & - \\
  			\toprule[0.5pt]
  			\toprule[1.0pt]
  	\end{tabular}}
  	\caption{\scriptsize{
  			Our CGAN predictions for the full width of the $ cc\bar{c}\bar{c} $ state (in units of MeV), based on A1 and A2, compared with the experimental data \cite{ParticleDataGroup:2022pth} and the theoretical estimates.}}
  	\label{tab:ccccwidth}
  \end{table}

\begin{table}
	\renewcommand{\arraystretch}{1.5}
	\renewcommand\tabcolsep{2.9pt}
	\scalebox{0.44}{
		\begin{tabular}{crl|cccc|ccc|ccc|ccc}
			\toprule[1.0pt]
			\toprule[0.5pt]
			\multicolumn{3}{c}{Systems} & \multicolumn{13}{c}{$bb\bar{b}\bar{b}$} \\
			\multicolumn{3}{l}{$J^{PC}$} &\multicolumn{2}{c}{ $0^{++}$} & $1^{++}$ & \multicolumn{1}{c}{$2^{++}$} & $0^{-+}$ & $1^{-+}$ & \multicolumn{1}{c}{$2^{-+}$} & $0^{+-}$ & $1^{+-}$ & \multicolumn{1}{c}{$2^{+-}$} & $0^{--}$ & $1^{--}$ & $2^{--}$ \\
			\toprule[0.5pt]
			&&Our result A1  &\multicolumn{2}{c}{ $19096 \pm 394 $} & $ 18322\pm {413}$ & $ 17990\pm {446}$ & $ 19062\pm {630}$ &$ 18720\pm {610}$ & $ 17881\pm {650}$ &  $ 18130\pm {394}$  & $ 17720\pm {413}$  & $ 17223\pm {446}$ & $19355\pm {499}$ & $ 18642\pm {543}$  & $ 17475\pm {474}$  \\
			&&Our result A2  &\multicolumn{2}{c}{$ 19250 \pm 357 $} &$ 18387 \pm 310 $ &$ 17800 \pm 228 $ &$ 19589 \pm 539 $ &$ 18806 \pm 368 $ &$ 17332 \pm 310 $&$ 19881 \pm 391 $ &$ 19145 \pm 339 $ &$ 17809 \pm 247 $ &$ 20181 \pm 468 $&$ 19238 \pm 405 $&$ 18340 \pm 301 $\\
			\Xcline{1-16}{0.01pt}
			\multicolumn{2}{c}{\multirow{3}*{\makecell[c]{The QCD sum\\ rule method}}}&Ref. \cite{Wang:2021mma}&18130&-&18140&18150&18450&18560&-&-&18140&-&18470&18460&-\\
			\Xcline{3-3}{0.01pt}
			&&Ref. \cite{Wang:2017jtz,Wang:2018poa}&\multicolumn{2}{c}{18840}&-&18850&-&-&-&-&18840&-&-&18890&-\\
			&&Ref. \cite{Agaev:2023wua}&\multicolumn{2}{c}{18540}&-&-&-&-&-&-&-&-&-&-&-\\
			\Xcline{1-16}{0.01pt}
			\multicolumn{2}{c}{\multirow{6}*{\makecell[c]{The nonrelativistic\\ constituent\\quark models}}}&Ref. \cite{An:2022qpt}&19352&19240&-&\multicolumn{1}{c|}{19328}&-&-&-&-&19304&-&-&-&-\\
			\Xcline{3-3}{0.01pt}
			&&\multirow{2}*{Ref. \cite{Wang:2019rdo}}&19215&19247&-&19249&-&-&-&-&19247&-&-&-&-\\
			&&&19243&19305&-&19325&-&-&-&-&19311&-&-&-&-\\
			\Xcline{3-3}{0.01pt}
			&&Ref. \cite{Liu:2019zuc}&19322&19338&-&19341&-&-&-&-&19329&-&-&-&-\\
			\Xcline{3-3}{0.01pt}
			&&Ref. \cite{Zhang:2022qtp}&19200&19235&-&19225&-&-&-&-&19216&-&-&-&-\\
			\Xcline{3-3}{0.01pt}
			\Xcline{1-16}{0.01pt}
			\multicolumn{2}{c}{\makecell[c]{The Bethe-\\ Salpeter equations}}&Ref.\cite{Li:2021ygk}&\multicolumn{2}{c}{19205}&-&19253&-&-&-&-&19221&-&-&-&-\\
			\Xcline{1-16}{0.01pt}
			\multicolumn{2}{c}{\multirow{2}*{\makecell[c]{The relativistic\\ quark model}}}&Ref. \cite{Faustov:2020qfm}&\multicolumn{2}{c}{19314}&-&19330&-&-&-&-&19320&-&-&-&-\\
			\Xcline{3-3}{0.01pt}
			&&Ref. \cite{Lu:2020cns}&19201&19255&-&19262&-&-&-&-&19251&-&-&-&-\\
			\Xcline{1-16}{0.01pt}
			\multicolumn{2}{c}{Monte Carlo method}&Ref. \cite{Gordillo:2020sgc}&\multicolumn{2}{c}{19199}&-&19289&-&-&-&-&19276&-&-&-&-\\
			\Xcline{1-16}{0.01pt}
			\multicolumn{2}{c}{\multirow{2}*{\makecell[c]{
						The diquark model}}}&Ref. \cite{Berezhnoy:2011xn}&\multicolumn{2}{c}{18754}&-&18916&-&-&-&-&18808&-&-&-&-\\
			\Xcline{3-3}{0.01pt}
			&&Ref. \cite{Mutuk:2021hmi}&\multicolumn{2}{c}{19666}&-&19680&-&-&-&-&19673&-&-&-&-\\
			\Xcline{1-16}{0.01pt}
			\multicolumn{2}{c}{\makecell[c]{An effective\\ potential model} }&Ref. \cite{Zhao:2020nwy}&19154&19226&-&19232&-&-&-&-&19214&-&-&-&-\\
			\Xcline{1-16}{0.01pt}
			\multicolumn{2}{c}{\makecell[c]{ Multiquark color\\ flux-tube model}}&Ref. \cite{Deng:2020iqw}&\multicolumn{2}{c}{19329}&-&19387&-&-&-&-&19373&-&-&-&-\\
			\Xcline{1-16}{0.01pt}
			\multicolumn{2}{c}{\multirow{5}*{\makecell[c]{The chromo-\\magnetic model}}}&Ref. \cite{Karliner:2016zzc}&\multicolumn{2}{c}{18826}&-&-&-&-&-&-&-&-&-&-&-\\
			\Xcline{3-3}{0.01pt}
			&&\multirow{2}*{Ref. \cite{Wu:2016vtq}}&20155&20275&-&20243&-&-&-&-&20212&-&-&-&-\\
			&&&18834&18954&-&18921&-&-&-&-&18890&-&-&-&-\\
			\Xcline{3-3}{0.01pt}
			&&Ref. \cite{Weng:2020jao}&18836&18981&-&19000&-&-&-&-&18969&-&-&-&-\\
			\Xcline{3-3}{0.01pt}
			&&Ref. \cite{Zhuang:2021pci}&18834&18953&-&18921&-&-&-&-&18890&-&-&-&-\\
			\toprule[0.5pt]
			\toprule[1.0pt]
	\end{tabular}}
	\caption{\scriptsize{Our CGAN predictions for the mass of the $bb\bar{b}\bar{b}$
			state (in units of MeV), based on A1 and A2, compared with the theoretical estimates.}}
	\label{tab:bbbbmass}
\end{table}

\begin{table} 
	\renewcommand{\arraystretch}{1.5}	
	\scalebox{0.49}{
		\begin{tabular}{crl|cccccccccccc}
			\toprule[1.0pt]
			\toprule[0.5pt]
			\multicolumn{3}{c}{Systems} & \multicolumn{12}{c}
			{$bb\bar{b}\bar{b}$} \\
			\multicolumn{3}{c}{$J^{P(C)}$} & $0^{++}$ & $1^{++}$ & \multicolumn{1}{c|}{$2^{++}$} & $0^{-+}$ & $1^{-+}$ & \multicolumn{1}{c|}{$2^{-+}$} & $0^{+-}$ & $1^{+-}$ & \multicolumn{1}{c|}{$2^{+-}$} & $0^{--}$ & $1^{--}$ & $2^{--}$ \\
			\toprule[0.5pt] && Our result A1  & $10.8\pm {4}$ & $7.9\pm {3}$ & $6.1\pm {3.7}$ & $10.5\pm {4}$ & $6.5\pm {3}$ & $5.5\pm {3}$ & $19\pm {8}$ & $13\pm {4}$ & $9\pm {3}$ & $18\pm {7}$& $10\pm {2}$ & $7.3\pm {3}$ \\
			&& Our result A2  &$6.6\pm {2.3}$ & $2.3\pm {0.7}$ &$1.4\pm {0.3}$  & $6.7\pm {1.2}$  & $1.9\pm {0.3}$&$1 \pm {0.2} $  & $7.1\pm {2.1}$  & $2.5\pm {0.3}$  & $2.1\pm {0.3}$ & $9\pm {1.2}$ & $4.2\pm {0.2}$ & $1.4\pm {0.1}$  \\
			\toprule[0.5pt]\multicolumn{2}{c}{\multirow{3}*{\makecell[c]{The QCD sum\\ rule method}}} & \multirow{2}*{Ref. \cite{Agaev:2023ara}}& 9.6 &- & -& -& -&- &- &- & -&- &- & - \\ 
			&&& 144 &- & -& -& -&- &- &- & -&- &- & - \\
			\Xcline{3-3}{0.01pt}&& Ref. \cite{Agaev:2023gaq}& 94 &-  & -& -& -&- &- &- & -&- &- & - \\
			\toprule[0.5pt]\multicolumn{2}{c}{\makecell[c]{The quasi-compact \\ diquark-antidiquark model }}& Ref. \cite{Esposito:2018cwh} & 0.001-10 &- & & -& -& -&- &- &- & -&- &- \\
			\toprule[0.5pt]\multicolumn{2}{c}{\makecell[c]{The QCD-String-junction\\ picture}}& Ref. \cite{Karliner:2016zzc} & 1.2 &- & & -& -& -&- &- &- & -&- &- \\
			\toprule[0.5pt]
			\toprule[1.0pt]
	\end{tabular}}
	\caption{\scriptsize{Our CGAN predictions for the full width of the $bb\bar{b}\bar{b}$
			state (in units of MeV), based on A1 and A2, compared with the theoretical estimates.}}
	\label{tab:bbbbwidth}
\end{table}

\subsection{The $ bb\bar{b}\bar{c} (cb\bar{b}\bar{b}) $ state}
Here, we will focus on the $ bb\bar{b}\bar{c} (cb\bar{b}\bar{b}) $ system.
Our CGAN predictions for the mass and width of this heavy tetraquark are shown in Tables \ref{tab:bbbcmass} and \ref{tab:bbbcwidth}.
We present the predicted results based on two approaches (A1 and A2) considering 
the quantum states of $J^{P} = 0^{+(-)}, 1^{+(-)}, 2^{+(-)}$, 
compared to the different theoretical models. 
As shown in Table \ref{tab:bbbcmass}, the CGAN predictions for the mass based on the both approaches 
fall within a similar range to those from other theoretical models. 
Now, let us turn our attention to Table \ref{tab:bbbcwidth}, 
which presents the CGAN predictions for the width of the $ bb\bar{b}\bar{c} (cb\bar{b}\bar{b}) $ state,
based on two approaches (A1 and A2) and compares them with theoretical models. 
For the state of $ 0^{+} $ there is a good agreement between our results and the QCD sum rule method \cite{Agaev:2024uza}. 
However, for the other quantum states, our predictions are notably higher than available theoretical results. 
It is possible that the width reported by some theoretical models does not represent the total decay width, 
as these models may not have included all relevant decay channels. 
In contrast, our CGAN model estimates the total decay width, 
which could explain the discrepancies observed between our predictions and those from other theoretical approaches.

\begin{table}
	\renewcommand{\arraystretch}{1.5}
	\renewcommand\tabcolsep{2.9pt}
	\scalebox{0.67}{
		\begin{tabular}{crl|cccc|cccc}
			\toprule[1.0pt]
			\toprule[0.5pt]
			\multicolumn{3}{c}{Systems} & \multicolumn{8}{c}{$bb\bar{b}\bar{c}\; (cb\bar{b}\bar{b})$} \\
			\multicolumn{3}{l}{$J^{PC}$}&\multicolumn{1}{c}{$0^{+}$}&\multicolumn{2}{c}{$1^{+}$}&$2^{+}$&\multicolumn{1}{c}{$0^{-}$}&\multicolumn{2}{c}{$1^{-}$}&$2^{-}$\\
			\toprule[0.5pt]
			&&Our result A1  &\multicolumn{1}{c}{$16113 \pm 550 $}&\multicolumn{2}{c}{$ 16031\pm {554}$}&$ 15667\pm {536}$ &\multicolumn{1}{c}{ $ 15888\pm {567}$}&\multicolumn{2}{c}{$ 15526\pm {561}$}& $ 15245\pm {508}$ \\
			&&Our result A2  &\multicolumn{1}{c}{$ 16318 \pm 491 $}&\multicolumn{2}{c}{$ 16161 \pm 418 $}&$ 15908 \pm 403 $ &\multicolumn{1}{c}{$ 16723 \pm 626 $ }&\multicolumn{2}{c}{$ 16264 \pm 503 $}&$ 15586 \pm 454 $  \\
			\Xcline{1-11}{0.01pt}
			\multicolumn{2}{c}{\makecell[c]{The QCD sum\\ rule method}}&Ref. \cite{Agaev:2024uza}&\multicolumn{1}{c}{15697}&\multicolumn{2}{c}{-}&-&-&\multicolumn{2}{c}{-}&-\\
			\Xcline{1-11}{0.01pt}
			\multicolumn{2}{c}{\multirow{5}*{\makecell[c]{The nonrelativistic\\ constituent\\quark models}}}&\multirow{2}*{Ref. \cite{An:2022qpt}}&16044&16043&16125&16149&-&\multicolumn{2}{c}{-}&-\\
			&&&16163&16144&-&-&-&\multicolumn{2}{c}{-}&-\\
			\Xcline{3-3}{0.01pt}
			&&\multirow{3}*{Ref. \cite{Liu:2019zuc}}&16190&\multicolumn{2}{c}{16167}&16176&-&\multicolumn{2}{c}{-}&-\\
			&&&16141&\multicolumn{2}{c}{16164}&-&-&\multicolumn{2}{c}{-}&-\\
			&&&-&\multicolumn{2}{c}{16148}&-&-&\multicolumn{2}{c}{-}&-\\
			\Xcline{3-3}{0.01pt}
			&&\multirow{2}*{Ref. \cite{Zhang:2022qtp}}&16061&16046&16079&16089&-&\multicolumn{2}{c}{-}&-\\
			&&&16100&16089&-&-&-&\multicolumn{2}{c}{-}&-\\
			\Xcline{1-11}{0.01pt}
			\multicolumn{2}{c}{The rel. quark model}&Ref. \cite{Faustov:2020qfm}&16109&\multicolumn{2}{c}{16117}&16132&-&\multicolumn{2}{c}{-}&-\\
			\Xcline{1-11}{0.01pt}
			\multicolumn{2}{c}{Monte Carlo method}&Ref. \cite{Gordillo:2020sgc}&16040&\multicolumn{2}{c}{16013}&16129&-&\multicolumn{2}{c}{-}&-\\
			\Xcline{1-11}{0.01pt}
			\multicolumn{2}{c}{\multirow{3}*{\makecell[c]{ Multiquark color\\ flux-tube model}}}&\multirow{3}*{Ref. \cite{Deng:2020iqw}}&\multicolumn{1}{c}{16158}&\multicolumn{2}{c}{16151}&16182&-&\multicolumn{2}{c}{-}&-\\
			&&&\multicolumn{1}{c}{16126}&\multicolumn{2}{c}{16230}&16274&-&\multicolumn{2}{c}{-}&-\\
			&&&\multicolumn{1}{c}{16175}&\multicolumn{2}{c}{16179}&-&-&\multicolumn{2}{c}{-}&-\\
			\Xcline{1-11}{0.01pt}
			\multicolumn{2}{c}{\multirow{7}*{\makecell[c]{The chromo-\\magnetic model}}}&\multirow{4}*{Ref. \cite{Wu:2016vtq}}&16952&16840&16884&16917\\
			&&&16823&16915&-&-&-&\multicolumn{2}{c}{-}&-\\
			&&&15713&15729&15773&15806\\
			&&&15841&15804&-&-&-&\multicolumn{2}{c}{-}&-\\
			\Xcline{3-3}{0.01pt}
			&&\multirow{3}*{Ref. \cite{Weng:2020jao}}&15712&15719&15851&15882\\
			&&&15862&15851&-&-&-&\multicolumn{2}{c}{-}&-\\
			&&&-&15854&-&-&-&\multicolumn{2}{c}{-}&-\\
			\toprule[0.5pt]
			\toprule[1.0pt]
	\end{tabular}}
	\caption{\scriptsize{Our CGAN predictions for the mass of the $bb\bar{b}\bar{c}\; (cb\bar{b}\bar{b})$
			state (in units of MeV), based on A1 and A2, compared with the theoretical estimates.}}
	\label{tab:bbbcmass}
\end{table}

\begin{table}
	\renewcommand{\arraystretch}{1.5}
	\centering
	\scalebox{0.73}{
		\begin{tabular}{crl|ccc|ccc}
			\toprule[1.0pt]
			\toprule[0.5pt]
			\multicolumn{3}{c}{Systems} & \multicolumn{6}{c}{$bb\bar{b}\bar{c}\; (cb\bar{b}\bar{b})$} \\
			\multicolumn{3}{c}{$J^{P(C)}$} & $0^{+}$ & $1^{+}$ & \multicolumn{1}{c|}{$2^{+}$} & $0^{-}$ & $1^{-}$ &  $2^{-}$\\
			\toprule[0.5pt]
			&&Our result A1 &$35\pm {12}$  &$11\pm {1.5}$ &$2.7\pm {0.3}$ &$98\pm {23}$& $10.5\pm {2.5}$&$5\pm {1.1}$  \\
			&&Our result A2 &$48\pm {11}$ & $20\pm {4}$ &$9.5\pm {1.1}$  &$42\pm {7}$ &$15\pm {0.9}$&$12 \pm {2} $ \\
			\Xcline{1-9}{0.01pt}
			\multicolumn{2}{c}{\makecell[c]{The chiral quark\\model}}& Ref. \cite{Hu:2022zdh}&6.1& 6.9& 8.5&- & -&- \\
			\Xcline{1-9}{0.01pt}
			\multicolumn{2}{c}{\makecell[c]{The quark potential\\model}}& Ref. \cite{Wu:2024hrv}&10 & 2& 14&- & -&- \\
			\Xcline{1-9}{0.01pt}
			\multicolumn{2}{c}{\makecell[c]{The QCD sum\\ rule method}}& Ref. \cite{Agaev:2024uza}&36 &-& -&- & -&- \\
			\toprule[0.5pt]
			\toprule[1.0pt]
	\end{tabular}}
	\caption{\scriptsize{Our CGAN predictions for the width of the $bb\bar{b}\bar{c}\; (cb\bar{b}\bar{b})$
			state (in units of MeV), based on A1 and A2, compared with the theoretical estimates.}}
	\label{tab:bbbcwidth}
\end{table}

\subsection{The $cc\bar{c}\bar{b}\; (bc\bar{c}\bar{c})$ state}  
The $cc\bar{c}\bar{b}\; (bc\bar{c}\bar{c})$ states represent intriguing tetraquark candidates. 
These exotic hadrons have garnered significant interest due to their potential implications for understanding the quark-gluon interactions and the structure of multi-quark states. 
Tables \ref{tab:cccbarbbarmass} and \ref{tab:cccbarbbarwidth} present our CGAN model predictions for the mass and decay width of these tetraquark states compared to those from other theoretical models.
In Table \ref{tab:cccbarbbarmass}, we summarize the mass predictions for the 
$cc\bar{c}\bar{b}\; (bc\bar{c}\bar{c})$ tetraquark state across different theoretical models. 
The table presents results for various quantum numbers $ J^{PC} $,
as predicted by our CGAN model (shown in two scenarios, A1 and A2), alongside results from several other approaches including QCD sum rules, lattice QCD, nonrelativistic constituent quark models, relativistic quark models, Monte Carlo methods, and others.
By comparing these predictions, we assess the consistency and potential deviations between our CGAN-based predictions and those from other methods, highlighting the strengths and unique aspects of our model.

In Table \ref{tab:cccbarbbarwidth}, we present the predicted widths for the $cc\bar{c}\bar{b}\; (bc\bar{c}\bar{c})$ tetraquark states, calculated using our CGAN model under two scenarios, A1 and A2. The table lists the widths for various quantum numbers $ J^{PC} $,
as predicted by our model, and we compare these values with those obtained from other theoretical methods, such as the chiral quark model, quark potential model, and QCD sum rule method. This comparison provides insight into the consistency and differences among  the results,
contributing to a better understanding of the decay properties of these exotic hadrons.
As noted earlier, the widths reported by some theoretical models may not reflect the total decay width due to incomplete consideration of all decay channels. 
This discrepancy is also evident in Table \ref{tab:cccbarbbarwidth}, 
where we compare our CGAN model's predictions for the total width of the
$cc\bar{c}\bar{b}\; (bc\bar{c}\bar{c})$ tetraquark states with those from other approaches. 
While our model accounts for the total decay width, some of the alternative models predict smaller widths, 
possibly due to the exclusion of certain decay channels. 
This highlights the potential advantages of our CGAN-based approach in estimating the complete decay width.

\begin{table}
	\renewcommand{\arraystretch}{1.5}
	\renewcommand\tabcolsep{2.9pt}
	\scalebox{0.65}{
		\begin{tabular}{crl|cccc|cccc}
			\toprule[1.0pt]
			\toprule[0.5pt]
			\multicolumn{3}{c}{Systems} & \multicolumn{8}{c}{$cc\bar{c}\bar{b}\; (bc\bar{c}\bar{c})$} \\
			\multicolumn{3}{l}{$J^{PC}$}&\multicolumn{1}{c}{$0^{+}$}&\multicolumn{2}{c}{$1^{+}$}&$2^{+}$&\multicolumn{1}{c}{$0^{-}$}&\multicolumn{2}{c}{$1^{-}$}&$2^{-}$\\
			\toprule[0.5pt]
			&&Our result A1  & $9630 \pm 340 $& \multicolumn{2}{c}{$ 9517\pm {318}$}  & $ 9108\pm {275}$ & $ 9421\pm {401}$ & \multicolumn{2}{c}{$ 9146\pm {333}$}& $ 8910\pm {298}$  \\ 
			&&Our result A2  &$ 9740 \pm 242 $ &\multicolumn{2}{c}{$ 9645 \pm 212 $ }&$ 9317 \pm 231 $ & $ 10278 \pm 214 $ &\multicolumn{2}{c}{ $ 9573 \pm 251 $} & $ 9066 \pm 260 $ \\
			\Xcline{1-11}{0.01pt}
			\multicolumn{2}{c}{\makecell[c]{The QCD sum\\ rule method}}&Ref. \cite{Agaev:2024uza}&9680&\multicolumn{2}{c}{-}&-&-&\multicolumn{2}{c}{-}&-\\
			\Xcline{1-11}{0.01pt}
			\multicolumn{2}{c}{\makecell[c]{Lattice-QCD}}&Ref. \cite{Yang:2021hrb}&9813&\multicolumn{2}{c}{9823}&10563&-&\multicolumn{2}{c}{-}&-\\
			\Xcline{1-11}{0.01pt}
			\multicolumn{2}{c}{\multirow{9}*{\makecell[c]{The nonrelativistic\\ constituent\\quark models}}}&\multirow{3}*{Ref. \cite{An:2022qpt}}&9621&\multicolumn{2}{c}{9625}&9731&-&\multicolumn{2}{c}{-}&-\\
			&&&9766&\multicolumn{2}{c}{9627}&-&-&\multicolumn{2}{c}{-}&-\\
			&&&-&\multicolumn{2}{c}{9706}&-&-&\multicolumn{2}{c}{-}&-\\
			\Xcline{3-3}{0.01pt}
			&&\multirow{3}*{Ref. \cite{Liu:2019zuc}}&9787&\multicolumn{2}{c}{9773}&9768&-&\multicolumn{2}{c}{-}&-\\
			&&&9715&\multicolumn{2}{c}{9752}&-&-&\multicolumn{2}{c}{-}&-\\
			&&&-&\multicolumn{2}{c}{9727}&-&-&\multicolumn{2}{c}{-}&-\\
			\Xcline{3-3}{0.01pt}
			&&\multirow{3}*{Ref. \cite{Zhang:2022qtp}}&9732&\multicolumn{2}{c}{9718}&9713&-&\multicolumn{2}{c}{-}&-\\
			&&&9665&\multicolumn{2}{c}{9699}&-&-&\multicolumn{2}{c}{-}&-\\
			&&&-&\multicolumn{2}{c}{9676}&-&-&\multicolumn{2}{c}{-}&-\\
			\Xcline{1-11}{0.01pt}
			\multicolumn{2}{c}{{\makecell[c]{The relativistic\\ quark model}}}&Ref. \cite{Faustov:2020qfm}&9572&9602&9619&9647&-&\multicolumn{2}{c}{-}&-\\
			\Xcline{1-11}{0.01pt}
			\multicolumn{2}{c}{Monte Carlo method}&Ref. \cite{Gordillo:2020sgc}&9615&\multicolumn{2}{c}{9610}&9719&-&\multicolumn{2}{c}{-}&-\\
			\Xcline{1-11}{0.01pt}
			\multicolumn{2}{c}{\multirow{3}*{\makecell[c]{ Multiquark color\\ flux-tube model}}}&\multirow{3}*{Ref. \cite{Deng:2020iqw}}&9314&\multicolumn{2}{c}{9343}&9443&-&\multicolumn{2}{c}{-}&-\\
			&&&9670&\multicolumn{2}{c}{9683}&9732&-&\multicolumn{2}{c}{-}&-\\
			&&&9753&\multicolumn{2}{c}{9766}&9839&-&\multicolumn{2}{c}{-}&-\\
			\Xcline{1-11}{0.01pt}
			\multicolumn{2}{c}{\multirow{5}*{\makecell[c]{The chromo-\\magnetic model}}}& \multirow{3}*{Ref. \cite{Wu:2016vtq}}&10144&10174&-&-&-&\multicolumn{2}{c}{-}&-\\
			&&&10322&10282&10273&-&-&\multicolumn{2}{c}{-}&-\\
			&&&-&10231&-&-&-&\multicolumn{2}{c}{-}&-\\
			\Xcline{3-3}{0.01pt}
			&& \multirow{2}*{Ref. \cite{Weng:2020jao}}&9318&9335&9484&9526&-&\multicolumn{2}{c}{-}&-\\
			&&&9506&9499&-&-&-&\multicolumn{2}{c}{-}&-\\
			\toprule[0.5pt]
			\toprule[1.0pt]
	\end{tabular}}
	\caption{\scriptsize{Our CGAN predictions for the mass of the $cc\bar{c}\bar{b}\; (bc\bar{c}\bar{c})$
			state (in units of MeV), based on A1 and A2, compared with the theoretical estimates.}}
	\label{tab:cccbarbbarmass}
\end{table}

\begin{table}
	\renewcommand{\arraystretch}{1.5}
	\centering
	\scalebox{0.7}{
		\begin{tabular}{crl|ccc|ccc}
			\toprule[1.0pt]
			\toprule[0.5pt]
			\multicolumn{3}{c}{Systems} & \multicolumn{6}{c}{$cc\bar{c}\bar{b}\; (bc\bar{c}\bar{c})$} \\
			\multicolumn{3}{c}{$J^{P(C)}$} & $0^{+}$ & $1^{+}$ & \multicolumn{1}{c|}{$2^{+}$} & $0^{-}$ & $1^{-}$ &  $2^{-}$\\
			\toprule[0.5pt]
			&&Our result A1 &$96\pm {23}$ &$ 26\pm {6}$ &$23\pm {7}$ & $25\pm {3}$& $14\pm {2.5}$&$8.8\pm {1.4}$  \\
			&&Our result A2  &$87\pm {12}$ &$ 32\pm {9}$ &$8.6\pm {0.9}$ & $103\pm {17}$& $29\pm {7}$&$12\pm {2}$  \\
			\Xcline{1-9}{0.01pt}
			\multicolumn{2}{c}{\makecell[c]{The chiral quark\\model}}& Ref. \cite{Hu:2022zdh}&8.4 & 7.2& 11.1&- & -&- \\
			\Xcline{1-9}{0.01pt}
			\multicolumn{2}{c}{\makecell[c]{The quark potential\\model}}& Ref. \cite{Wu:2024hrv}&4 & 1& 20&- & -&- \\
			\Xcline{1-9}{0.01pt}
			\multicolumn{2}{c}{\makecell[c]{The QCD sum\\ rule method}}& Ref. \cite{Agaev:2024uza}&54.7 &-& -&- & -&- \\
			\toprule[0.5pt]
			\toprule[1.0pt]
	\end{tabular}}
	\caption{\scriptsize{Our CGAN predictions for the full width of the $cc\bar{c}\bar{b}\; (bc\bar{c}\bar{c})$
			state (in units of MeV), based on A1 and A2, compared with the theoretical estimates.}}
	\label{tab:cccbarbbarwidth}
\end{table}

\subsection{The $cc\bar{b}\bar{b}$ ($bb\bar{c}\bar{c}$) state} 
The $cc\bar{b}\bar{b}$ ($bb\bar{c}\bar{c}$) state  is a fully-heavy tetraquark with an antisymmetric wavefunction, 
where both the diquarks and antiquarks exhibit antisymmetry.
Tables \ref{tab:ccbbmass} and \ref{tab:ccbbwidth} illustrate the predictions from our CGAN model, 
based on the A1 and A2 approaches, for the mass and decay width of this fully-heavy tetraquark, 
compared with various theoretical models.
Upon examining the predicted mass values in Table \ref{tab:ccbbmass}, 
it is evident that the CGAN model predictions, based on A1 and A2, 
generally fall within the range calculated by the theoretical models, 
with a slight tendency to predict slightly higher values.
Let us now turn our discussion to the width predictions in Table \ref{tab:ccbbwidth}. The CGAN model results, 
represented by A1 and A2, show a noticeable variation in the predicted decay widths. For the $cc\bar{b}\bar{b}$ ($bb\bar{c}\bar{c}$) system, 
the results from A1 generally give smaller decay widths compared to those from A2. 
For example, the decay widths for the $0^{+}$ and $1^{+}$ states predicted by A1 are significantly lower than those predicted by A2. 
The widths predicted by A1, for instance, range from $7.6 \pm 0.8 \,~\text{MeV}$ to $80 \pm 18 \,~\text{MeV}$, 
while those predicted by A2 range from $52 \pm 19 \,~\text{MeV}$ to $149 \pm 29 \,~\text{MeV}$.  
When compared with the theoretical models listed in the table, 
the CGAN results (both A1 and A2) are in reasonable agreement with the QCD sum rule method and the other theoretical approaches. 
For instance, the QCD sum rule method, as predicted in Refs. \cite{Agaev:2023tzi} and \cite{Agaev:2024pej}, 
shows a wide range of predictions, with some values closer to the lower or higher end of the spectrum from the CGAN model. 
This suggests that while the CGAN model’s predictions show some variation, 
they fall within the range of values predicted by established methods.
The differences between the CGAN model predictions (A1 and A2) and
the QCD sum rule results might be attributed to the varying approximations used in the different theoretical models. 
These comparisons provide valuable insights into the predictive power of the CGAN model, 
suggesting that it offers results consistent with established theoretical frameworks while also highlighting the sensitivity of decay widths to model assumptions and parameter choices.

\begin{table}
	\renewcommand{\arraystretch}{1.5}
	\renewcommand\tabcolsep{2.9pt}
	\scalebox{0.65}{
		\begin{tabular}{crl|cccc|cccc}
			\toprule[1.0pt]
			\toprule[0.5pt]
			\multicolumn{3}{c}{Systems} & \multicolumn{7}{c}{$cc\bar{b}\bar{b}$ ($bb\bar{c}\bar{c}$)} \\
			\multicolumn{3}{l}{$J^{PC}$}&\multicolumn{1}{c}{$0^{+}$}&\multicolumn{2}{c}{$1^{+}$}&$2^{+}$&$0^{-}$&$1^{-}$&$2^{-}$\\
			\toprule[0.5pt]
			&&Our result A1  &\multicolumn{2}{c}{$13233 \pm 623 $} & $ 12977\pm {556}$ & $ 12414\pm {428}$ & $ 12849\pm {589}$ &   $ 12497\pm {561}$  & $ 11557\pm {518}$ \\ 
			&&Our result A2 &\multicolumn{2}{c}{$ 13303 \pm 481 $}&$ 13043 \pm 504 $& $ 12751 \pm 465 $ &$ 13120 \pm 787 $ &$ 12943 \pm 611 $ &$ 12727 \pm 629 $  \\
			\Xcline{1-10}{0.01pt}
			\multicolumn{2}{c}{\multirow{2}*{\makecell[c]{The QCD sum\\ rule method}}}&Ref. \cite{Agaev:2024xdc}&-&&-&-&13092&13150&-\\
			\Xcline{3-3}{0.01pt}
			&&Ref. \cite{Agaev:2024pil}&-&&-&12795&-&-&-\\
			\Xcline{3-3}{0.01pt}
			&&\multirow{2}*{Ref.  \cite{Agaev:2023tzi}}&12715&&-&-&-&-&-\\
			&&&13383&&-&-&-&-&-\\
			\Xcline{3-3}{0.01pt}
			&&\multirow{2}*{Ref.  \cite{Agaev:2024pej}}&-&&12715&-&-&-&-\\
			&&&-&&13383&-&-&-&-\\
			\Xcline{1-10}{0.01pt}
			\multicolumn{2}{c}{\multirow{5}*{\makecell[c]{The nonrelativistic\\ constituent\\quark models}}}&Ref. \cite{An:2022qpt}&12920&13008&12940&12961&-&-&-\\
			\Xcline{3-3}{0.01pt}
			&&\multirow{2}*{Ref. \cite{Wang:2019rdo}}&12847&12866&12864&12868&-&-&-\\
			&&&12886&12946&12924&12940&-&-&-\\
			\Xcline{3-3}{0.01pt}
			&&Ref. \cite{Liu:2019zuc}&12947&13039&12960&12972&-&-&-\\
			\Xcline{3-3}{0.01pt}
			&&Ref. \cite{Zhang:2022qtp}&12880&12981&12890&12902&-&-&-\\
			\Xcline{1-10}{0.01pt}
			\multicolumn{2}{c}{{\makecell[c]{The relativistic\\ quark model}}}&Ref. \cite{Faustov:2020qfm}&\multicolumn{2}{c}{12846}&12859&12883&-&-&-\\
			\Xcline{1-10}{0.01pt}
			\multicolumn{2}{c}{Monte Carlo method}&Ref. \cite{Gordillo:2020sgc}&\multicolumn{2}{c}{12865}&12908&12926&-&-&-\\
			\Xcline{1-10}{0.01pt}
			\multicolumn{2}{c}{\multirow{2}*{\makecell[c]{
						The diquark model}}}&\multirow{2}*{Ref. \cite{Mutuk:2022nkw}}&\multicolumn{2}{c}{12401}&12409&12427&-&-&-\\
			&&&\multicolumn{2}{c}{12381}&12390&12408&-&-&-\\
			\Xcline{1-10}{0.01pt}
			\multicolumn{2}{c}{\multirow{4}*{\makecell[c]{ Multiquark color\\ flux-tube model}}}&\multirow{4}*{Ref. \cite{Deng:2020iqw}}&12597&12938&12660&12695&-&-&-\\
			&&&12906&13023&12945&12960&-&-&-\\
			&&&12940&12986&13024&13041&-&-&-\\
			&&&12963&-&-&-&-&-&-\\
			\Xcline{1-10}{0.01pt}
			\multicolumn{2}{c}{\multirow{3}*{\makecell[c]{The chromo-\\magnetic model}}}& \multirow{2}*{Ref. \cite{Wu:2016vtq}}&13496&13634&13560&13595&-&-&-\\
			&&&12597&12734&12660&12695&-&-&-\\
			\Xcline{3-3}{0.01pt}
			&&Ref. \cite{Weng:2020jao}&12596&12712&12672&12703&-&-&-\\
			\Xcline{1-10}{0.01pt}
			\multicolumn{2}{c}{\makecell[c]{The nonrelativistic \\ chiral quark model}}&Ref. \cite{Chen:2020lgj}&
			\multicolumn{2}{c}{12684}&12737&12791&-&-&-\\
			\toprule[0.5pt]
			\toprule[1.0pt]
	\end{tabular}}
	\caption{\scriptsize{Our CGAN predictions for the mass of the $cc\bar{b}\bar{b}$ ($bb\bar{c}\bar{c}$)
			state (in units of MeV), based on A1 and A2, compared with the theoretical estimates.}}
	\label{tab:ccbbmass}
\end{table}

\begin{table}
	\renewcommand{\arraystretch}{1.5}
	\centering
	\scalebox{0.7}{
		\begin{tabular}{crl|ccc|ccc}
			\toprule[1.0pt]
			\toprule[0.5pt]
			\multicolumn{3}{c}{Systems} & \multicolumn{6}{c}{$cc\bar{b}\bar{b}$ ($bb\bar{c}\bar{c}$)} \\
			\multicolumn{3}{c}{$J^{P(C)}$} & $0^{+}$ & $1^{+}$ & \multicolumn{1}{c|}{$2^{+}$} & $0^{-}$ & $1^{-}$ &  $2^{-}$\\
			\toprule[0.5pt]
			&&Our result A1  &$77\pm {15}$ &$ 23\pm {3}$ &$11\pm {1.4}$ & $80\pm {18}$& $16\pm {2.7}$&$7.6\pm {0.8}$  \\
			&&Our result A2 &$149\pm {29}$ &$ 52\pm {13}$ &$42\pm {9}$ & $118\pm {36}$& $59\pm {12}$&$52\pm {19}$  \\
			\Xcline{1-9}{0.01pt}
			\multicolumn{2}{c}{\multirow{6}*{\makecell[c]{The QCD sum\\ rule method}}}& \multirow{2}*{Ref. \cite{Agaev:2023tzi}}&63 &- & -& -& -&-\\
			&&&79 &- & -& -& -&-  \\
			\Xcline{3-3}{0.01pt}
			&& Ref. \cite{Agaev:2024pej}&44.3&- & -& -& -&- \\
			&&&82.5&- & -& -& -&- \\
			\Xcline{3-3}{0.01pt}
			&& Ref. \cite{Agaev:2024pil} &- &- &55.5 & -& -& -\\
			\Xcline{3-3}{0.01pt}
			&& Ref. \cite{Agaev:2024xdc} &- &- &- & 63.7& 53.5& -\\
			\toprule[0.5pt]
			\toprule[1.0pt]
	\end{tabular}}
	\caption{\scriptsize{Our CGAN predictions for the full width of the $cc\bar{b}\bar{b}$ ($bb\bar{c}\bar{c}$)
			state (in units of MeV), based on A1 and A2, compared with the theoretical estimates.}}
	\label{tab:ccbbwidth}
\end{table}

\subsection{The $cb\bar{c}\bar{b}$ state}
\begin{figure}[h!]	
	\begin{center} 	
		\includegraphics[width=1.1\textwidth]{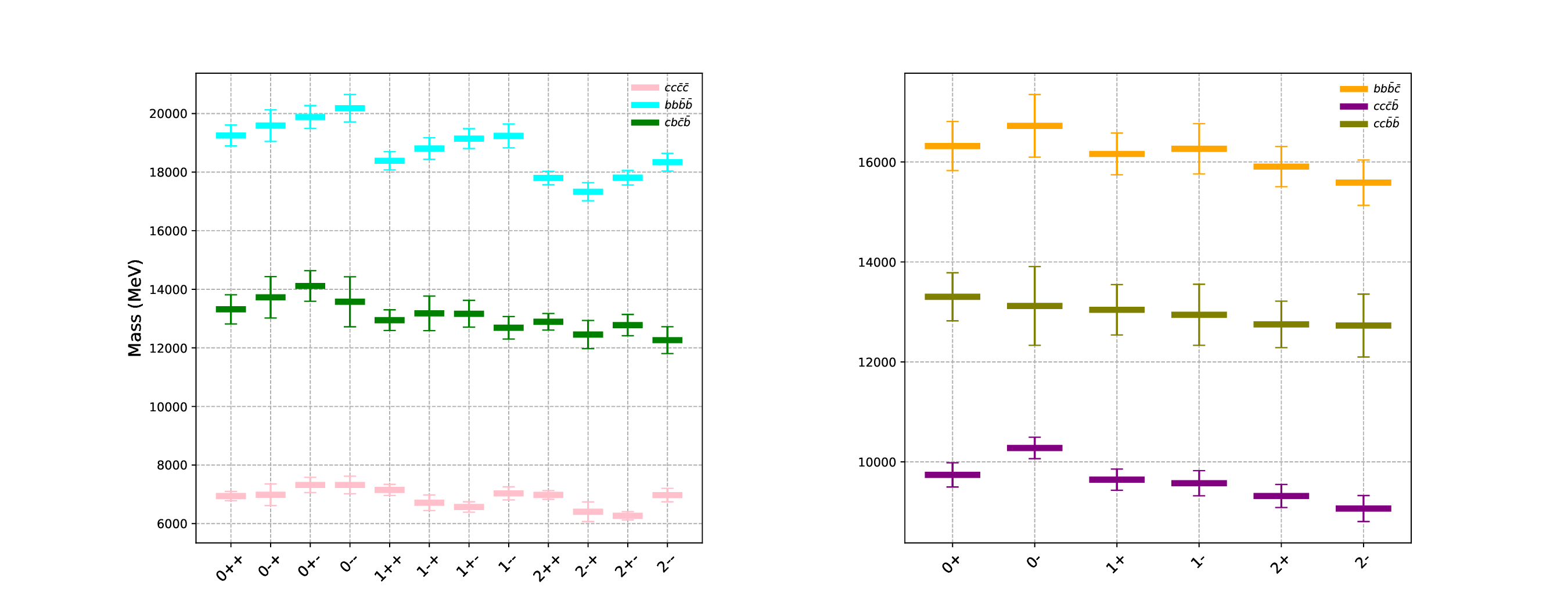}	
	\end{center}	
	\caption{\scriptsize{Predicted mass spectra of the fully-heavy tetraquark states by the CGAN model based on \textbf{A2} for various quantum states. 
			The left panel shows the predicted mass spectra for the $cc\bar{c}\bar{c}$, $bb\bar{b}\bar{b}$, and $cb\bar{c}\bar{b}$systems across different quantum states ($J^{PC}$). 
			The right panel displays the predicted mass spectra for $bb\bar{b}\bar{c}$, $cc\bar{c}\bar{b}$, and $cc\bar{b}\bar{b}$ systems. }}	\label{A2}
\end{figure}

Finally, we examine the $cb\bar{c}\bar{b}$ system. 
This system is  neutral and has a certain C-parity. 
Table \ref{tab:cbcbarbbarmass} presents mass predictions of the CGAN model (A1 and A2) 
for the $cb\bar{c}\bar{b}$ system across various quantum states ($J^{PC}$), 
compared to multiple theoretical models. 
It is obvious that the predicted results of the CGAN model based on A2 tend to be slightly higher compared to A1, 
indicating that the change in dataset features between these two approaches influences the predictions. 
However, the CGAN results, based on both approaches, are in good agreement with the theoretical model results, 
with only minor discrepancies.
Table \ref{tab:cbcbarbbarwidth} illustrates the width predictions for the $cb\bar{c}\bar{b}$ state from our CGAN model (A1 and A2) 
in comparison with the QCD sum rule method. 
The width predictions from the CGAN model based on A2 are generally higher than those from A1. 
When compared to the available QCD sum rule results, the CGAN predictions show a trend of being smaller than the QCD sum rule values.
For instance, the mass spectra of fully-heavy tetraquarks predicted by the CGAN model based on A2 for various quantum states are shown in Fig. \ref{A2}, demonstrating the capabilities of this method in exploring heavy tetraquark systems.
It is worth noting that, unlike traditional quark models where the mass hierarchy typically increases with total angular momentum $J$, our predictions focus on reproducing the experimentally observed mass hierarchy and determining which angular
momentum value leads to a mass prediction that most closely matches the experimental result (specifically for $cc\bar{c}\bar{c}$) or presents a more reasonable result (for other systems), rather than examining a spectrum of states with identical quark content, where a systematic increase in mass with spin is typically observed. In other words, our approach focuses on identifying the most probable spin configuration corresponding to the known mass, even if the predicted hierarchy does not strictly follow conventional expectations.

\begin{table} 
	\renewcommand{\arraystretch}{1.5} 
	\renewcommand\tabcolsep{2.9pt}	
	\scalebox{0.45}{
		\begin{tabular}{crl|cccc|ccc|ccc|ccc}
			\toprule[1.0pt]\toprule[0.5pt]
			\multicolumn{3}{c}{Systems} & \multicolumn{13}{c}{$cb\bar{c}\bar{b}$} \\
			\multicolumn{3}{l}{$J^{PC}$} & \multicolumn{2}{c}{ $0^{++}$} & $1^{++}$ & \multicolumn{1}{c}{$2^{++}$} & $0^{-+}$ & $1^{-+}$ & \multicolumn{1}{c}{$2^{-+}$} & $0^{+-}$ & $1^{+-}$ & \multicolumn{1}{c}{$2^{+-}$} & $0^{--}$ & $1^{--}$ & $2^{--}$ \\
			\toprule[0.5pt]&&Our result A1  &  \multicolumn{2}{c}{$12924 \pm 478 $}  & $ 12810\pm {376}$  & $ 12276\pm {329}$ & $ 13994\pm {680}$ & $ 12574\pm {469}$  & $ 12598\pm {426}$ & $ 12178\pm {468}$ & $ 11982\pm {421}$& $ 11515\pm {405}$ & $12854\pm {627}$  & $ 11829\pm {483}$ & $ 11185\pm {467}$ \\
			&& Our result A2  & \multicolumn{2}{c}{$ 13316 \pm 498 $ } & $ 12947 \pm 353 $ & $ 12891 \pm 283 $ &$ 13727 \pm 707 $ &$ 13178 \pm 590 $ & $ 12453 \pm 481 $&$ 14113 \pm 523 $ & $ 13165 \pm 458 $ &$ 12778 \pm 363 $&$ 13574 \pm 854 $&$ 12686 \pm 384 $ &$ 12264 \pm 459 $ \\
			\Xcline{1-16}{0.01pt}\multicolumn{2}{c}{\makecell[c]{The QCD sum\\ rule method}} & Ref. \cite{Agaev:2024wvp}& \multicolumn{2}{c}{12697}&-&-&-&-&-&-&-&-&-&-&-\\
			\Xcline{1-16}{0.01pt}\multicolumn{2}{c}{\multirow{3}*{\makecell[c]{Lattice-QCD}}} & \multirow{3}*{Ref. \cite{Yang:2021hrb}}&\multicolumn{2}{c}{12820}&12858&12826&-&-&-&-&12858&-&-&-&-\\ &&& \multicolumn{2}{c}{13449}&13002&13321&-&-&-&-&13002&-&-&-&-\\&&&\multicolumn{2}{c}{-}&13290&-&-&-&-&-&13290&-&-&-&-\\
			\Xcline{1-16}{0.01pt}\multicolumn{2}{c}{\multirow{6}*{\makecell[c]{The nonrelativistic\\ constituent \\
						quark models}}}& \multirow{2}*{Ref. \cite{An:2022qpt}}&12760&12851&12857&12882&-&-&-&-&12797\quad 12852&-&-&-&-\\
			&&&12989&13009&12960&12971&-&-&-&-&12999\quad 13056&-&-&-&-\\
			\Xcline{3-3}{0.01pt}&&\multirow{2}*{Ref. \cite{Liu:2019zuc}}&12854&12974&12933&12933&-&-&-&-&12881\quad 12909&-&-&-&-\\
			&&&12931&13024&12992&13021&-&-&-&-&13004\quad 13020&-&-&-&-\\
			\Xcline{3-3}{0.01pt}&&\multirow{2}*{Ref. \cite{Zhang:2022qtp}}&12783&12850&12851&12852&-&-&-&-&12802\quad 12835&-&-&-&-\\
			&&&12966&13035&12938&12964&-&-&-&-&12949\quad 12964&-&-&-&-\\
			\Xcline{1-16}{0.01pt}\multicolumn{2}{c}{{\makecell[c]{The relativistic\\ quark model}}}&Ref. \cite{Faustov:2020qfm}&12813&12824&12831&12849&-&-&-&-&12826 \quad 12831&-&-&-&-\\
			\Xcline{1-16}{0.01pt}\multicolumn{2}{c}{Monte Carlo method}&Ref. \cite{Gordillo:2020sgc} & \multicolumn{2}{c}{12534}&12569&12582&-&-&-&-&12510&-&-&-&-\\
			\toprule[0.5pt]
			\toprule[1.0pt]
	\end{tabular}}
	\caption{\scriptsize{Our CGAN predictions for the mass of the $cb\bar{c}\bar{b}$
			state (in units of MeV), based on A1 and A2, compared with the theoretical estimates.}}	
	\label{tab:cbcbarbbarmass}
\end{table}

\begin{table}
	\renewcommand{\arraystretch}{1.5}
	\centering
	\scalebox{0.53}{
		\begin{tabular}{crl|cccccccccccc}
			\toprule[1.0pt]
			\toprule[0.5pt]
			\multicolumn{3}{c}{Systems} & \multicolumn{12}{c}{$cb\bar{c}\bar{b}$} \\
			\multicolumn{3}{c}{$J^{P(C)}$} & $0^{++}$ & $1^{++}$ & \multicolumn{1}{c|}{$2^{++}$} & $0^{-+}$ & $1^{-+}$ & \multicolumn{1}{c|}{$2^{-+}$} & $0^{+-}$ & $1^{+-}$ & \multicolumn{1}{c|}{$2^{+-}$} & $0^{--}$ & $1^{--}$ & $2^{--}$ \\
			\toprule[0.5pt]
			&&Our result A1 &$54\pm {9}$ &$14\pm {3}$ &$1.9\pm {0.4}$ & $64\pm {8}$& $7.4\pm {0.5}$&$1.5\pm {0.4}$ &$5.3\pm {0.9}$ & $6.1\pm {1}$&$0.8\pm {0.2}$ & $46\pm {7}$& $11\pm {4}$& $1\pm {0.2}$ \\
			&&Our result A2&$131\pm {34}$ & $88\pm {32}$ &$37\pm {12}$  &$113\pm {33}$  &$44\pm {11}$&$ 24\pm {8} $  &$153\pm {22}$  & $80\pm {15}$  & $30\pm {9}$ & $165\pm {35}$ & $69\pm {21}$ &$20\pm {4}$  \\
			\toprule[0.5pt]
			\multicolumn{2}{c}{\multirow{3}*{\makecell[c]{The QCD sum\\ rule method}}}& Ref. \cite{Agaev:2024wvp}&142.4 &- & -& -& -&- &- &- & -&- &- & - \\
			\Xcline{3-3}{0.01pt}
			&& Ref. \cite{Agaev:2024mng}&- &104.2 & -& -& -&- &- &- & -&- &- & - \\
			\Xcline{3-3}{0.01pt}
			&& Ref. \cite{Agaev:2024qbh} &- &- &117.4 & -& -& -&- &- &- & -&- &- \\
			\toprule[0.5pt]
			\toprule[1.0pt]
	\end{tabular}}
	\caption{\scriptsize{Our CGAN predictions for the full width of the $cb\bar{c}\bar{b}$
			state (in units of MeV), based on A1 and A2, compared with the theoretical estimates.}}
	\label{tab:cbcbarbbarwidth}
\end{table}

\section {Summary and conclusion}\label{SC}
Fully-heavy tetraquarks are exotic hadrons composed entirely of heavy quarks (\( c \) and \( b \)). 
They are particularly interesting because their unique properties stem from the large mass of the quarks involved. 
The large mass of heavy quarks simplifies the understanding of interactions, 
resulting in non-relativistic behavior that can be effectively described using potential models or non-relativistic approximations. 
Besides, fully-heavy tetraquarks are expected to have relatively narrow widths, 
as their decay channels are constrained by the significant mass difference between the tetraquark and its decay products. 
These states provide valuable insights into the dynamics of strong interactions and the involvement of heavy quarks in exotic hadrons. 
Consequently, the study of fully-heavy tetraquarks enhances our understanding of quarkonium and multiquark physics.
Although experimental data on fully-heavy tetraquarks is lacking, 
the study of these intriguing states remains an active area of theoretical research. 
Various theoretical models have been employed to explore these heavy systems and estimate their mass and decay width.
In this study, we developed CGAN frameworks to predict the mass and total decay width of fully-heavy tetraquarks, 
utilizing two distinct approaches for preparing and classifying the dataset (A1 and A2). 
We presented the numerical results from our CGAN models for the mass and total decay width of fully-heavy tetraquarks, 
comparing them to the available experimental data  and theoretical predictions. The CGAN model's predictions align well with existing data, 
demonstrating its reliability in predicting both the mass and width of fully-heavy tetraquarks. 
This consistency across models increases confidence that the CGAN approach produces realistic and valuable results. 
Despite limited experimental evidence, this study suggests that, alongside theoretical models, 
the ML techniques including the CGAN can provide valuable predictions to guide future searches for fully-heavy tetraquarks. 
In fact, the CGAN approach could become a strong contender for future studies of heavy tetraquark systems, 
complementing existing theoretical models to provide more precise results.
Our results are expected to provide useful insights for experiments, 
offering information that may not be easily obtained from most existing theoretical models for exotic hadrons.

\section*{ACKNOWLEDGEMENTS}
S.~R. and M.~M. would like to express their heartfelt gratitude to the organizers of the MITP Summer School on "Machine Learning in Particle Theory" for their invaluable support, insightful lectures, and the opportunity to engage with cutting-edge advancements in the intersection of machine learning and particle physics. S.~R., M.~M., and K.~A are grateful to the CERN-TH division for their warm hospitality.


\begin{thebibliography}{0}
\bibitem{Zweig:1964jf}
	G.~Zweig,
	``An SU(3) model for strong interaction symmetry and its breaking. Version 2,''
\end{thebibliography}
\end{document}